\documentclass[12pt]{article}

\usepackage{amsmath,amssymb,amsthm}
\usepackage{bbm}
\usepackage[unicode]{hyperref}
\usepackage{graphicx}
\usepackage{geometry}
\usepackage{fancyhdr}

\newcommand{\cev}[1]{\reflectbox{\ensuremath{\vec{\reflectbox{\ensuremath{#1}}}}}}
\newcommand{\tr}{\operatorname{tr}}
\newcommand{\str}{\operatorname{str}}
\newcommand{\sdet}{\operatorname{sdet}}
\newcommand{\p}{\partial}
\newcommand{\abs}[1]{\left|#1\right|}

\title{\LARGE\bf K\"ahler-Einstein and \\
 K\"ahler scalar flat supermanifolds}

\author{J.~P. Ang$^{a}$\footnote{\href{mailto:jianpeng.ang@stonybrook.edu}{jianpeng.ang@stonybrook.edu}}\, ,  Martin Ro\v{c}ek$^{a}$\footnote{\href{mailto:martin.rocek@stonybrook.edu}{martin.rocek@stonybrook.edu}}\, 
and John Schulman$^{b}$\footnote{\href{joschu@berkeley.edu}{joschu@berkeley.edu}}
\bigskip \\
\normalsize \emph{$^{a}$C. N. Yang Institute for Theoretical Physics}\\
\normalsize \emph{State University of New York, Stony Brook, NY 11790-3840} \\
\\
\normalsize \emph{$^{b}$Department of Electrical Engineering and Computer Science} \\ 
\normalsize \emph{University of California, Berkeley CA}
}

\date{}

\begin{document}

\maketitle
\thispagestyle{fancy}\rhead{YITP-SB-16-19}

\begin{abstract}
Two results regarding K\"ahler supermanifolds with potential $K=A+C\theta\bar\theta$ are shown. First, if the supermanifold is K\"ahler-Einstein, then its base (the supermanifold of one lower fermionic dimension and with K\"ahler potential $A$) has constant scalar curvature. As a corollary, every constant scalar curvature K\"ahler supermanifold has a unique superextension to a K\"ahler-Einstein supermanifold of one higher fermionic dimension. Second, if the supermanifold is {\em itself} scalar flat, then its base satisfies the equation
\begin{equation}
\phi^{\bar ji}\phi_{i\bar j}=2\Delta_0 S_0 + R_0^{\bar ji}R_{0i\bar j} - S_0^2~,
\end{equation}
where $\Delta_0$ is the Laplace operator, $S_0$ is the scalar curvature, and $R_{0i\bar j}$ is the Ricci tensor of the base, and $\phi$ is some harmonic section on the base. Remarkably, precisely this equation arises in the construction of certain supergravity compactifications. Examples of bosonic manifolds satisfying the equation above are discussed. 
\end{abstract}

\vfill
%



\section{Introduction}
In \cite{Rocek:2004bi} it was shown that a super-Ricci flat K\"ahler supermanifold of complex fermionic dimension $1$ has a K\"ahler scalar flat bosonic base. The result is actually slightly more general -- it applies to any supermanifold, of possibly higher fermionic dimension, with a super-K\"ahler potential of the form 
$K=A+C\theta\bar\theta$, where the bosonic base is replaced by a supermanifold of one fermionic dimension lower with super-K\"ahler potential $A$. In this article, we continue the study of supermanifolds with potential $K=A+C\theta\bar\theta$.

In section 2, we find that K\"ahler-Einstein supermanifolds have a base (super)manifold that is K\"ahler with constant scalar curvature (cscK). Equivalently, any cscK (super)manifold with (super)K\"ahler potential $A$ has a unique extension to a K\"ahler supermanifold of one higher fermionic dimension with potential $A+C\theta\bar\theta$. As an example, we look at the cscK manifold $\mathbb{CP}^{n}$, whose extension is the K\"ahler-Einstein supermanifold $\mathbb{CP}^{n|1}$.

In section 3, we find that if the supermanifold is scalar flat, the base supermanifold satisfies a constraint on its curvature,
\begin{equation}
\phi^{\bar ji}\phi_{i\bar j}=2\Delta_0 S_0 + R_0^{\bar ji}R_{0i\bar j} - S_0^2,
\label{eqn.master2}
\end{equation}
where $\Delta_0$ is the Laplace operator, $S_0$ is the scalar curvature, and $R_{0i\bar j}$ is the Ricci tensor of the base, and $\phi$ is some harmonic section on the base. Remarkably, this equation is satisfied by certain symplectic leaves of solutions of IIB supergravity with $AdS_3$ factors and of $d=11$ supergravity with $AdS_2$ factors \cite{Kim:2005ez,Kim:2006qu}. We review the method introduced in \cite{Gauntlett:2006ns,Gauntlett:2007ts} of generating bosonic manifolds satisfying equation \eqref{eqn.master2} from a positively curved K\"ahler-Einstein manifold, and use this method to construct superscalar flat K\"ahler examples.

\section{K\"ahler-Einstein supermanifolds}

In this section, we show that if a K\"ahler supermanifold of complex dimension $(D_0|D_1)$ (complex bosonic dimension $D_0$ and fermionic dimension $D_1$) 
and super-K\"ahler potential\footnote{In the case of one fermionic dimension, this is the most general form of the K\"ahler potential if we do not allow the potential to depend on fermionic parameters.}
\begin{equation}
K = A+\theta\bar\theta C \label{eqn.kpot}
\end{equation}
satisfies Einstein's equations, then the base (super)manifold obtained by setting $\theta=0$ must have constant (super)scalar curvature. Note that the base (super)manifold has dimension $(D_0|D_1-1)$ and is (super)K\"ahler with potential $A$.

We define\footnote{Following \cite{Rocek:2004bi}, we use a convention where holomorphic derivatives act from the left and antiholomorphic derivatives act from the right. It is related to a convention where all derivatives act from the left by
\begin{equation}
\vec\p_I X\cev\p_{\bar J} = (-)^{\abs{\bar J}(\abs{X}+1)}\vec\p_I\vec\p_{\bar J}X~.
\end{equation}}
\begin{equation}\label{eqn.ric}      \
\begin{array}{rclrcl}
     g_{I\bar J}& =& \vec\p_I K \cev\p_{\bar J}~,  &\sqrt{g}&=&\sdet g_{I\bar J}~, \\[2mm]
      R_{I\bar J}&=&-\vec\p_I(\log\sqrt{g})\cev\p_{\bar J}~, ~~~& S &= &\str g^{\bar JI}R_{I\bar K}~, \\
   \end{array}
\end{equation}
where $\str,\sdet$ are the usual supertrace and superdeterminant; for any supermatrix 
\begin{equation}
X:=\left(   \begin{matrix} 
      A & B \\
      C & D \\
   \end{matrix}\right)~,~
\end{equation}
we have
\begin{equation}
\str X= \tr A-\tr D~,~~\sdet X = \frac{\det A}{\det (D-CA^{-1} B)}= \frac{\det (A-BD^{-1}C)}{\det D}~.
\end{equation}
The normalization of the super-Ricci scalar $S$ differs from the standard definition by a factor of 2. 

Einstein's equation is
\begin{equation}
R_{I\bar J}-Sg_{I\bar J}+\Lambda g_{I\bar J}=0~.
\end{equation}
Taking the supertrace yields
\begin{equation}
S+(-S+\Lambda)(D_0-D_1) = 0~,
\end{equation}
where we have used that $\str g^{\bar JI}g_{I\bar K}=\str \delta^J_K=D_0-D_1$ is the (complex) superdimension. Therefore, Einstein supermanifolds have constant super Ricci scalar, given by
\begin{equation}
S = \frac{\Lambda(D_0-D_1)}{D_0-D_1-1}~.
\end{equation}
Einstein's equation may be rewritten as
\begin{equation}
R_{I\bar J}=\Upsilon g_{I\bar J}~, \label{eqn.tracerev}
\end{equation}
where $\Upsilon=\Lambda/(D_0-D_1-1)$. We note that in complex dimension $D_0-D_1=1$, this trace reversal does not work, 
so we will \emph{define} a superdimension 1 Einstein supermanifold by \eqref{eqn.tracerev}. Using \eqref{eqn.ric}, we can rewrite (\ref{eqn.tracerev}) as:
\begin{equation}
\vec\p_I(\log\sqrt{g}+\Upsilon K)\cev\p_J = 0~.
\end{equation}
This implies that we can perform a holomorphic coordinate transformation so that in the new coordinates, we have
\begin{equation}
\sqrt{g} = e^{-\Upsilon K}~.
\end{equation}
Substituting this in \eqref{eqn.kpot} and using $\theta^2=0$, we find:
\begin{equation}
\sqrt{g}=e^{-\Upsilon A}(1-\theta\bar\theta\Upsilon C)~. \label{eqn.cond}
\end{equation}

We now find the conditions imposed on the base manifold by the super Einstein equation. First, we find that
\begin{equation}
g_{I\bar J}=\vec\p_I K \cev\p_{\bar J} = \begin{pmatrix} A_{i\bar j}+\theta\bar\theta C_{i\bar j} & C_i\theta \\ \bar\theta C_{\bar j} & C \end{pmatrix}~,
\end{equation}
so
\begin{align}
\sqrt{g} =& \frac{1}{C}\det \left(A_{i\bar j}+\theta\bar\theta C_{i\bar j}-C_i\theta\frac{1}{C}\bar\theta C_{\bar j}\right) \nonumber \\
=& \frac{1}{C}\det A_{i\bar j}\left(1+\theta\bar\theta C\Delta_0\log C\right), \label{eqn.sqrtg}
\end{align}
where subscripts on $A$ and $C$ denote derivatives, $A^{\bar ji}$ is the inverse of the metric on the base manifold $A_{i\bar j}$, and $\Delta_0$ is (half) the Laplacian on the base manifold\footnote{When the base is a supermanifold, we use the definition
of the super-Laplacian given below (\ref{superdel}), and the $\det$'s in (\ref{eqn.sqrtg}) become $\sdet$'s. } $\Delta_0=g^{\bar ji}\p_i\p_{\bar j}$. Comparing \eqref{eqn.sqrtg} to \eqref{eqn.cond} yields
\begin{align}
e^{-\Upsilon A} &= \frac{1}{C}\det A_{i\bar j}, \label{eqn.res1} \\
-e^{-\Upsilon A}\Upsilon C &= \det A_{i\bar j}\ \Delta_0\log C. \label{eqn.res2}
\end{align}
In particular, we find that
\begin{align}
-\Upsilon =& \Delta_0\log C = \Delta_0\log(e^{\Upsilon A}\det A_{i\bar j}) \nonumber\\
=& \Delta_0(\Upsilon A+\log\det A_{i\bar j}) \nonumber\\
=& \Upsilon (D_0-D_1+1) - S_0~,
\end{align}
where $S_0$ is the Ricci scalar of the base manifold $S_0=-\Delta_0\log\det A_{i\bar j}$. 

We therefore find that the base manifold is cscK with
\begin{equation}
S_0 = \Upsilon(D_0-D_1+2) = \Lambda\frac{D_0-D_1+2}{D_0-D_1-1}~.
\end{equation}

We can also carry out these steps in the opposite direction, obtaining a fermionic extension of any cscK manifold. In the following example, we find the extension of $\mathbb{CP}^n$.

\subsection{Superextension of $\mathbb{CP}^{n}$}
$\mathbb{CP}^{n}$ with the Fubini-Study metric is a K\"ahler manifold given by the potential
\begin{equation}
A = \log(1+z\bar z) := \log(1+\delta_{k\bar k}z^k\bar z^{\bar k})~.
\end{equation}
The metric is
\begin{equation}
A_{i\bar j} = \frac{\delta_{i\bar j}}{1+z\bar z}-\frac{z_i\bar z_{\bar j}}{(1+z\bar z)^2}~,
\end{equation}
and we can compute
\begin{equation}
\log\det A_{i\bar j} = \log(1+z\bar z)^{-n-1} = -(n+1)A~,
\end{equation}
which shows that $\mathbb{CP}^{n}$ is K\"ahler-Einstein with $\Upsilon_0=n+1$, and therefore is cscK with $S_0=n(n+1)$.

Equation \eqref{eqn.res1} then tells us that the superextension of $\mathbb{CP}^{n}$ has
\begin{equation}
C = e^{\Upsilon A}\det A_{i\bar j}~.
\end{equation}
Substituting this into \eqref{eqn.res2} yields
\begin{align}
-\Upsilon =& \Delta_0\log(e^{\Upsilon A}\det A_{i\bar j}) \\
=& (\Upsilon-(n+1))n~,
\end{align}
where we have used $\Delta_0A=n$. This shows that $\Upsilon=n$, so the $D_1=1$ superextension of $\mathbb{CP}^{n}$ has potential
\begin{equation}
K = \log(1+z\bar z)+\theta\bar\theta(1+z\bar z)^{-1} = \log(1+z\bar z+\theta\bar\theta)~.
\end{equation}
This space is the super complex projective space $\mathbb{CP}^{n|1}$.

\section{Super K\"ahler scalar flat}
We now study the constraints that super-scalar flatness imposes on the base manifold. Using such a supermanifold as a base allows us to use the method
described in the previous section to find super K\"ahler-Einstein supermanifolds.

First, we define the super-Laplacian by
\begin{align}\label{superdel}
\Delta =& \str g^{\bar JI}\vec\p_I\cev\p_{\bar K} \nonumber \\
=& g^{\bar ji}\p_i\p_{\bar j}+g^{\bar j\theta}\vec\p_\theta\p_{\bar j}-
g^{\bar\theta i}\p_i\cev\p_{\bar\theta}-g^{\bar\theta\theta}\vec\p_\theta\cev\p_{\bar\theta}~.
\end{align}
Notice that $\Delta K=D_0-D_1$ is the superdimension, and $-\Delta\log\sdet\sqrt{g}=S$ is the super scalar curvature.

We compute the inverse metric\footnote{The inverse of a supermatrix in block form is 
\begin{equation}
\begin{pmatrix} A & B \\ C & D \end{pmatrix} = \begin{pmatrix} (A-BD^{-1}C)^{-1} & -(A-BD^{-1}C)^{-1}BD^{-1} \\ -D^{-1}C(A-BD^{-1}C)^{-1} & (D-CA^{-1}B)^{-1} \end{pmatrix}~.
\end{equation}}
\begin{align}
g^{\bar JI} =& \begin{pmatrix} (A_{i\bar j}+\theta\bar\theta(\log C)_{i\bar j})^{-1} & -A^{\bar ji}(\log C)_i\theta \\ -\bar\theta(\log C)_{\bar j}A^{\bar ji} & (C+\theta\bar\theta C_{\bar j}A^{\bar ji}C_i)^{-1} \end{pmatrix} \nonumber \\
=& \begin{pmatrix} A^{\bar ji} - \theta\bar\theta C(\log C)^{\bar ji} & -(\log C)^{\bar j}\theta \\ -\bar\theta(\log C)^i & C^{-1}-\theta\bar\theta(\log C)^i(\log C)_i \end{pmatrix}~,
\end{align}
where in the last line we have used the base metric $A_{i\bar j}$ to raise and lower bosonic indices. Using the earlier result \eqref{eqn.sqrtg}
\begin{align}
\log\sdet\sqrt{g} =& \log\left(\frac{1}{C}\det A_{i\bar j}(1+\theta\bar\theta C\Delta_0\log C)\right) \nonumber\\
=& -\log C+\log\det A_{i\bar j}+\theta\bar\theta C\Delta_0\log C~,
\end{align}
we find that the condition for super-scalar flatness is
\begin{align}
0 =& \Delta \log\sdet\sqrt{g} \nonumber\\
=& \Delta_0(\log\det A_{i\bar j}-2\log C) \nonumber\\
&+\theta\bar\theta C\Big((\Delta_0\log C)^2+\Delta_0\Delta_0\log C+(\log C)^{\bar ji}((\log C)_{i\bar j}+R_{0i\bar j})\Big)
\end{align}
The bosonic component tells us that
\begin{equation}
\phi = 2\log C-\log\det A_{i\bar j} \label{eqn.phi}
\end{equation}
is a harmonic section on the base manifold ({\it i.e.,} $\Delta_0\phi=0$).
We use this relation to eliminate $\log C$ in the nilpotent component in favor of $\phi$, yielding the geometric identity
\begin{equation}
0 = (S_0)^2-2\Delta_0 S_0 + \phi^{\bar ji}\phi_{i\bar j} - R^{\bar ji}_0R_{0i\bar j}~.
\label{eqn.masterphi}
\end{equation}

\subsection{Global properties of $\phi$}
In one fermionic dimension, the most general holomorphic coordinate transformation is $z'^i = f^i(z),\theta' = h(z)\theta$, whose Jacobian is
\begin{equation}
\frac{\p(z',\theta')}{\p(z,\theta)} = \begin{pmatrix} \p_j f^i & 0 \\ \p_jh & h \end{pmatrix}~.
\end{equation}
This shows that holomorphic coordinate transformations are automatically split in one fermionic dimension, so the supermanifold is the whole space of a Grassmann-valued holomorphic line bundle $F$ defined by the transition function $h$.

The superdeterminant $\sqrt{g}$ transforms as $\sqrt{g'}=\sqrt{g}\abs{\det \p_j f^i/h}^{-2}$. The volume element of the base metric and $C$ transform as
\begin{align}
\det A_{i\bar j}'&=\det A_{i\bar j}\abs{\det \p_jf^i}^{-2}, \nonumber \\
C'&=C\abs{h}^2~.
\end{align}
Therefore, according to \eqref{eqn.phi}, $e^{\phi}$ is a section of the line bundle $D\otimes F^2\otimes \bar F^2$, where $D$ is the determinant bundle of the cotangent bundle of the base manifold. Its transition function is $e^{\phi'}=e^{\phi}\abs{\det \p_j f^i}^{2}\abs{h}^4$.

\subsection{Examples with constant $\phi$}
The simplest examples are products of K\"ahler-Einstein (super)manifolds of dimensions $(D_0^{(\alpha)}|D_1^{(\alpha)})$ with metric $g_{i\bar j}^{(\alpha)}$ and Ricci tensors satisfying $R^{(\alpha)}_{i\bar j}=\Upsilon^{(\alpha)}g_{i\bar j}^{(\alpha)}$, with locally constant\footnote{$\phi$ is necessarily locally constant in the case where the line bundle is trivial and the base manifold is bosonic and compact, since $\int\left|\nabla\phi\right|^2=-\int \phi\Delta_0\phi=0$ by partial integration.} $\phi$, so that the resulting equation
\begin{equation}
(S_0)^2-2\Delta_0S_0-R^{\bar ji}_0R_{0i\bar j} = 0.
\label{eqn.master}
\end{equation}
becomes an algebraic equation
\begin{equation}
\Big(\sum_\alpha \Upsilon^{(\alpha)}(D_0^{(\alpha)}-D_1^{(\alpha)})\Big)^2 - \sum_\alpha \left(\Upsilon^{(\alpha)}\right)^2(D_0^{(\alpha)}-D_1^{(\alpha)})=0~.
\end{equation}
Examples of solutions are $\mathbb{CP}^{n|n}$, $\mathbb{CP}^{n+1|n}$ and $\mathbb{CP}^1\times\mathbb{CP}^1\times H$, where $H$ is the hyperbolic plane. The superextensions of the first two spaces are respectively $\mathbb{CP}^{n|n+1}$ and $\mathbb{CP}^{n+1|n+1}$. Note that $\mathbb{CP}^{n|n}$ is both K\"ahler-Einstein with nonzero cosmological constant \emph{and} scalar flat (this is not a contradiction since $D_0-D_1=0$), while $\mathbb{CP}^{n|n+1}$ is Ricci flat \cite{Rocek:2004ha,Sethi:1994ch}.

Remarkably, \eqref{eqn.master} is also satisfied by symplectic leaves of certain supersymmetric solutions of IIB supergravity with an $AdS_3$ factor \cite{Kim:2005ez} or 11 dimensional supergravity with an $AdS_2$ factor \cite{Kim:2006qu}.\footnote{Equation \eqref{eqn.master} first appeared in \cite{Cariglia:2004kk}, for $D_0=2$, in the context of $U(1)$ and $SU(2)$ gauged supergravities in six dimensions.} In \cite{Gauntlett:2006ns,Gauntlett:2007ts}, the authors found a method of generating manifolds of dimension $D_0$ satisfying \eqref{eqn.master}, generalizing the supergravity solutions, starting from a (bosonic) positively curved K\"ahler-Einstein manifold of dimension $D_0-1$. In the following, we briefly review the method.

Let $d\hat s^2$ and $\hat\omega$ be respectively the metric and K\"ahler form of a $D_0-1$ dimensional K\"ahler-Einstein manifold with positive cosmological constant, normalized so that its scalar curvature is $\hat S_0=2D_0(D_0-1)$. Consider the following ansatz for the metric and Hermitian form of the $D_0$ dimensional manifold,
\begin{equation}
ds^2_0 = \frac{1}{x}\left(\frac{dx^2}{4x^2U} + U(d\psi+B)^2 + d\hat s^2\right)~,
\label{eqn.egmetric}
\end{equation}
\begin{equation}
\omega_0 = \frac{1}{x}\left(-\frac{1}{2x}dx\wedge(d\psi+B)+\hat\omega\right)~,
\label{eqn.egform}
\end{equation}
where $B$ satisfies $dB=2\hat\omega$, and $U=U(x)$ is an as yet undetermined function of $x$. It is straightforward to verify that $\omega_0$ is closed. The Einstein property of the $D_0-1$ dimensional K\"ahler manifold then ensures that the complex structure defined by $\omega_0$ is integrable. Therefore, the $D_0$ dimensional manifold is indeed K\"ahler, and its potential may be computed to be
\begin{equation}
A = \int^x \frac{dx'}{2{x'}^2U(x')}~.
\label{eqn.egpot}
\end{equation}
It can be explicitly checked that, for this ansatz, equation \eqref{eqn.master} reduces to a fourth order differential equation for $U(x)$. It is then straightforward to find polynomial solutions of $U(x)$ by inspection, see \cite{Gauntlett:2006ns} for details. We shall focus on a particular family of solutions
\begin{equation}
U = 1-\alpha x^{D_0-2}(x-1)^2~,
\end{equation}
where $\alpha$ is any constant.

We now determine the ranges of the coordinates giving rise to complete and nonsingular manifolds. Note that $U$ should be everywhere nonnegative to ensure positive definiteness of the metric, and that there is a finite distance curvature singularity at $x=\infty$. This means that we should exclude $\infty$ from the range of $x$, and hence we should take $\alpha$ positive.

We first discuss the case when $D_0>2$. For $\alpha<\alpha_0=D_0^{D_0}/4(D_0-2)^{D_0-2}$, $U$ has just one positive zero $x_0$. For $\alpha>\alpha_0$, $U$ has three positive zeroes, $x_0,x_1,x_2$ (ordered increasingly). Therefore, there are two ranges we could choose for $x$: $0<x\leq x_0$, or $x_1\leq x\leq x_2$. We discuss each in turn.
\begin{itemize}
\item
$0<x\leq x_0$: Note that $x\to 0$ is at an infinite distance away, so this is a noncompact space. Near $x=x_0$, $U$ depends linearly on $x-x_0$. We therefore approximate $U$ by $U'_0(x-x_0)$, and introduce the change of coordinates $\rho=-{U_0'}^{-1/2}x_0^{-3/2}(x-x_0)^{1/2}$, so the metric near $x_0$ becomes
\begin{equation}
ds^2 = d\rho^2 + \rho^2 {U_0'}^2 x_0^2 (d\psi+B)^2 + \frac{1}{x_0}d\hat s^2~.
\end{equation}
If we choose $\psi$ to have period $-2\pi/U'_0x_0$, then $x=x_0$ is just a polar coordinate singularity, so this manifold is smooth and noncompact. Its volume is infinite.
\item
$x_1\leq x\leq x_2$: This is a compact space, assuming that the $D_0-1$ K\"ahler-Einstein manifold is compact. By the same argument as above, near $x_1$ and $x_2$, the regions look like polar coordinate singularities, but in this case, it is not possible to choose a period for $\psi$ avoiding conical singularities at both $x_1$ and $x_2$. This is therefore a compact conifold.
\end{itemize}
If $\alpha=\alpha_0$, so that two zeroes meet, $x_0=x_1$, in either range $0<x<x_0$ or $x_1<x<x_2$, the region near $x_0=x_1$ is an infinitely long spike of finite volume.

For $D_0=2$, for $\alpha<1$ there is just one positive root, yielding the first scenario; while for $\alpha>1$ there are two positive roots, yielding the second scenario.

The scalar flat superextensions of these manifolds are given by
\begin{equation}
C^2=k\det A_{i\bar j}=\frac{k}{2}x^{-D_0-1}\det\hat \omega_{i\bar j},
\end{equation}
where $k$ is some positive constant.

\subsection{Examples with nontrivial $\phi$}
The supergravity solutions of \cite{Kim:2005ez,Kim:2006qu} were generalized in \cite{Donos:2008ug} to support magnetic fluxes. The fluxes enter the geometry by modifying \eqref{eqn.master} to include the flux action. For instance, the $D_0=4$ dimensional manifold giving rise to the $11$-dimensional supergravity solution supports a $(2,2)$ primitive field strength $F_{ij\bar k\bar l}$, which modifies the equation satisfied by the manifold to
\begin{equation}
(S_0)^2 - 2\Delta_0S_0 - R_0^{\bar ji}R_{0i\bar j} + F^{\bar l\bar kji}F_{ij\bar k\bar l} = 0~ .
\end{equation}
We adapt this method to generate a 2-form flux which would precisely give examples of manifolds satisfying \eqref{eqn.masterphi} with nontrivial $\phi_{i\bar j}$. Starting with the same ansatz \eqref{eqn.egmetric}, \eqref{eqn.egform} as before, but this time also include a 2-form flux
\begin{equation}
F = q~ d\left(x^{D_0-1}(d\psi+B)\right).
\end{equation}
It can be checked that $F$ is of type $(1,1)$, harmonic and primitive, so that it may indeed be written as the complex hessian of a local potential function $F_{i\bar j}=\phi_{i\bar j}$. Equation \eqref{eqn.masterphi} reduces, as before, to a fourth order differential equation for $U(x)$. We shall focus on a particular family of solutions
\begin{equation}
U = 1-\alpha x^{D_0-2}(\alpha^{-1} x-1)(\alpha^{-1} x-\beta)~ ,
\end{equation}
where $\beta=1+q/\sqrt{2}$, and $\alpha$ is an arbitrary constant we take to be positive to avoid the curvature singularity at $x=\infty$. The properties of $U$ are similar to the fluxless case; $U$ has one positive zero for small $\alpha$ and three for large $\alpha$. We can let $x$ take values from $x=0$ to the first positive zero of $U$, in which case we obtain an infinite volume smooth manifold; or between the two largest roots of $U$ when it has three positive zeroes, in which case we obtain a compact conifold.

The scalar flat superextensions of these manifolds are given by
\begin{equation}
C^2=e^\phi\det A_{i\bar j}=\frac{1}{2}e^\phi x^{-n-1}\det\hat \omega_{i\bar j},
\end{equation}
where $\phi$ is the local potential for the 2-form flux.

\section{Conclusion}
We have studied the relation between a K\"ahler supermanifold, with potential $K=A+C\theta\bar\theta$, and its base supermanifold, in the cases when the supermanifold is Einstein and when it is scalar flat. In the first case, we find that the base is cscK, and in the second case, the geometry of the base satisfies a differential constraint \eqref{eqn.master2}.  As described above, this differential constraint has been studied in the context of supergravity solutions; why the same equation arises in apparently unrelated contexts is unclear. Certainly the supermanifold construction is more concise, and it could lead to new insights.

Another direction for further study is to generalize the K\"ahler potential to $K=A+B\bar\theta+\theta\bar B+C\theta\bar\theta$, which is not related to our case by holomorphic coordinate transformations.

\section*{Acknowledgements}
Most of this work is based on John Schulman's 2005 Intel Science Talent Search project. MR and JPA are supported in part by NSF Grant No. PHY-1316617. We would like to thank Eoin \'O Colg\'ain for useful discussions and bringing reference \cite{Donos:2008ug} to our attention.

\bibliography{ric}
\bibliographystyle{jhep_td}

\end{document}